\documentclass[journal=jacsat,manuscript=article]{achemso}
\usepackage[table]{xcolor}
\usepackage{colortbl}
\usepackage{booktabs}
\definecolor{softblue}{RGB}{220,230,245}
\definecolor{softpink}{RGB}{245,220,230}

\usepackage[version=3]{mhchem}
\usepackage{amssymb}
\usepackage{graphicx}
\usepackage[colorlinks=true,breaklinks=true,linkcolor=black,citecolor=green,urlcolor=magenta]{hyperref}
\usepackage{physics}

\usepackage[table]{xcolor}

\definecolor{PastelBlue}{rgb}{0.68,0.85,0.90}
\definecolor{PastelPink}{rgb}{1.0,0.91,0.95}

\title{I-QMapper: Error-Aware Layout Optimization and Device Diagnostics for NISQ Hardware}

\author{Milana Bazayeva}
\affiliation{Center for Computational Life Sciences, Lerner Research Institute, The Cleveland Clinic, Cleveland, Ohio 44106, United States}

\author{Kenneth M. Merz Jr.}
\email{kmerz1@gmail.com}
\affiliation{Center for Computational Life Sciences, Lerner Research Institute, The Cleveland Clinic, Cleveland, Ohio 44106, United States}
\alsoaffiliation{Department of Chemistry, Michigan State University, East Lansing, Michigan 48824, United States}

\begin{document}
\maketitle

\begin{abstract}
Achieving high-fidelity execution on noisy intermediate-scale quantum (NISQ) hardware requires careful selection of physical qubit layouts, as gate errors, readout errors, and coherence times vary across the device and drift over time. Currently, qubit mapping is performed either through manual inspection of device calibration data or through automated layout pipelines, neither of which provides integrated, interactive layout visualization combined with calibration analytics. In this work, we present the Interactive Quantum Mapper (I-QMapper), a Jupyter-based, open-source tool for noise-aware layout selection, visualization, and analysis on superconducting quantum hardware. I-QMapper offers two operating modes: a general-purpose mode for arbitrary circuits, and a dedicated mode for quantum-chemistry applications, specifically tailored to the Local Unitary Cluster Jastrow (LUCJ) ansatz. Within each mode, a Design panel supports interactive layout construction, while an Error panel provides calibration analytics through four temporal viewing modes (Live, Snapshot, Intraday, and Multi-day range) together with threshold filtering and delta-mode comparison for drift identification. Each layout receives a Layout-Quality Score (LQS) that aggregates the readout and two-qubit gate errors of the layout into a single quality value. Starting from the automatic LUCJ circuit-generation provided by IBM Quantum, we extend it to a multi-programming setting in which multiple circuits are mapped onto a single quantum processing unit (QPU). I-QMapper further supports side-by-side visualization of two quantum backends and layout comparison, and session export for experimental reproducibility. By combining interactive exploration with calibration analytics, I-QMapper aims to support both rapid layout prototyping and informed noise-aware experimental design on NISQ devices.
\end{abstract}

Quantum computing is rapidly expanding across many fields, spanning from machine learning~\cite{Biamonte2017} to quantum chemistry~\cite{MottaPromiseToQuantum,QuantumProtein,BazayevaQuantumAFE,kaliakin2025implicit,shajan2024towards,wang2025samplebasedquantumdiagonalizationparallel,robledo2024chemistry}. However, current quantum devices remain in the noisy intermediate-scale quantum (NISQ) regime, where limited qubit counts, finite coherence times, cross-talk, and gate and readout errors set a strict bound on the depth and accuracy of executable circuits.

Within this regime, the choice of which physical qubits to use for a given logical circuit has a direct impact on output quality. Error rates vary by up to an order of magnitude across the qubits of a single device~\cite{tannu2019not}, and individual qubit and gate calibrations drift over hours due to periodic recalibration and environmental fluctuations. As a consequence, even small changes in layout selection can substantially alter circuit-execution outcomes~\cite{nation2023suppressing,paler2019influence}, making informed qubit selection a non-trivial optimization problem. This effect is particularly pronounced for structured ansätze such as those used in variational quantum eigensolvers (VQE), where the two-qubit gate topology of the circuit must match a specific subgraph of the device connectivity; the Local Unitary Cluster Jastrow (LUCJ) ansatz~\cite{Motta2023LUCJ}, central to this work, is a representative case.

Several approaches address layout selection and routing in current compilation pipelines. SABRE~\cite{li2019tackling} and its recent version LightSABRE~\cite{zou2024lightsabre} perform initial-layout selection and SWAP insertion to minimize routing overhead, optimizing for circuit depth and gate count rather than error rates, and are therefore calibration-agnostic by default. Other methods explicitly account for backend calibration: MQT QMAP~\cite{wille2023mqt} computes optimal qubit mappings under several cost functions, including error-aware ones. Q-fid~\cite{mao2025qfid} uses LSTM models to predict the fidelity of a transpiled circuit before execution. Finally, mapomatic~\cite{nation2023suppressing} works after transpilation, searching for alternative sets of physical qubits with the same connectivity and ranking them through a product of per-operation fidelities derived from calibration data.

The IBM Quantum platform itself exposes calibration data through a color-coded device topology, and Qiskit provides static views via \verb|plot_coupling_map| and \verb|plot_error_map|. These representations are useful as a first glance, but are static and difficult to track across the daily calibration updates. All of these approaches share a common assumption: the user has already produced or delegates to a transpiler an initial layout. What is missing is an environment that supports the layout design stage itself, with interactive visualization, manual or assisted construction, and live access to calibration analytics.

In this context, we developed I-QMapper, an open-source, Jupyter-based, calibration-aware qubit layout tool that fills this upstream gap. I-QMapper provides: i) two operating modes, the General ansatz for arbitrary circuits and a dedicated mode targeting the LUCJ ansatz for quantum-chemistry applications; ii) automatic LUCJ circuit generation, with support for the multi-programming regime; iii) a Layout-Quality Score (LQS) that quantifies the noise level of any selected layout based on readout and two-qubit gate errors; iv) temporal calibration inspection across snapshots spanning hours to days, including a delta mode that highlights drift in device parameters; v) a multi-vendor architecture, demonstrated through full IBM Quantum integration and extensible by design to additional providers.

The resulting layout can be exported and passed as \verb|initial_layout| to standard Qiskit compilation, optionally followed by post-selection ranking via mapomatic, making I-QMapper complementary to the existing tooling ecosystem rather than competing with it.

The remainder of this paper is organized as follows. Section \nameref{sec:architecture} describes the software architecture of I-QMapper. Section \nameref{sec:design} details the qubit-design workflow, including the Layout-Quality Score and LUCJ-aware layout construction. Section \nameref{sec:errors} presents the error-mode visualization and the time-resolved calibration inspection. 

\section{Software Architecture}
\label{sec:architecture}
I-QMapper is implemented in Python within the Jupyter ecosystem. The Jupyter ecosystem was chosen because it is the standard environment for Qiskit-based workflows, allowing I-QMapper to live alongside the user's circuit and execution code in the same notebook. The graphical user interface is built on \texttt{ipywidgets} \cite{widgets}, while the interactive visualization of the device topology and error map is rendered via Plotly \cite{plotly}. Communication with the IBM Quantum backends and retrieval of calibration data are handled through \texttt{qiskit-ibm-runtime} \cite{javadi2024quantum}. Additional dependencies include \texttt{rustworkx} \cite{rustworkx} for graph operations, \texttt{numpy} for numerical handling, and Kaleido/imageio for PNG and GIF export. The code base is organized into modules covering backend communication, data retrieval and caching, layout construction, error analysis, and UI support.

\begin{figure}[H]
    \centering
    \includegraphics[width=1\linewidth]{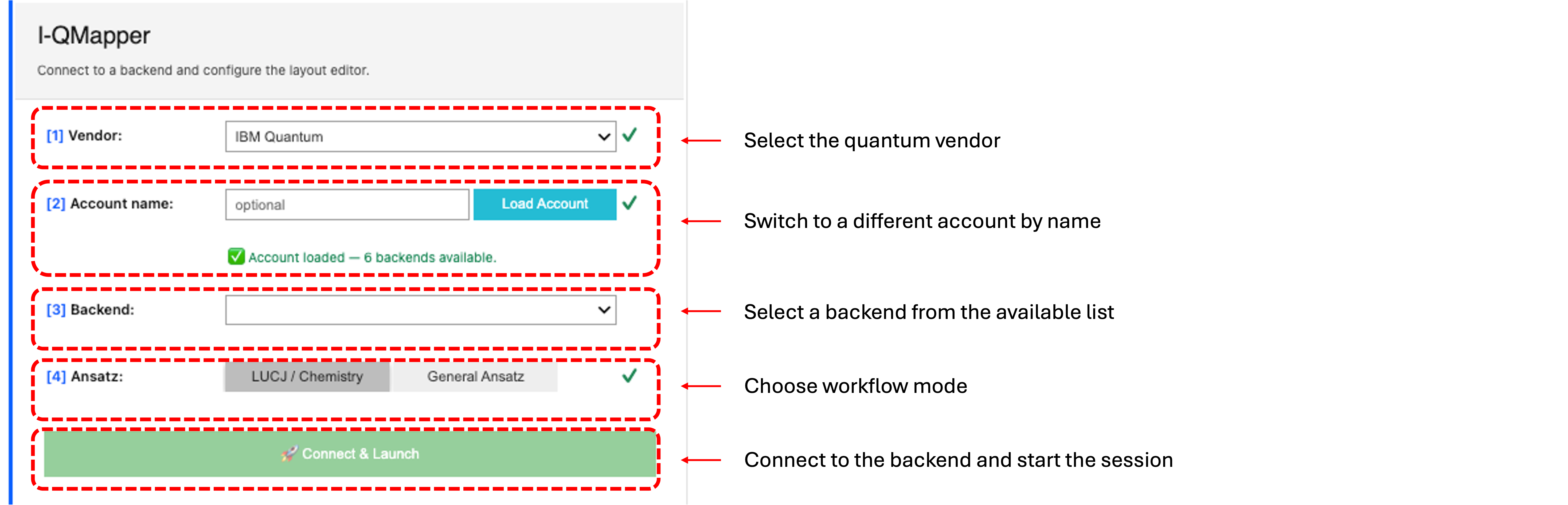}
\caption{The connection panel guides the user through five steps: i) selection of the quantum vendor; ii) optional specification of a saved Qiskit account name; iii) selection of a backend from the auto-populated list; iv) choice between the LUCJ or the General ansatz workflow; and v) launch of the mapper. The account configuration is handled outside of the I-QMapper via Qiskit.}
    \label{fig:Figure1}
\end{figure}

IBM Quantum authentication is intentionally decoupled from the tool: the user establishes the connection to the IBM Cloud beforehand via Qiskit's \texttt{save\_account} function, which stores the credentials locally, keeping credential management under full user control. The connection interface supports multi-account configurations through the optional \texttt{name} argument of \texttt{save\_account}, allowing the user to register and switch between several saved profiles, as shown in Figure~\ref{fig:Figure1}.

Backend calibration data (readout errors, two-qubit gate errors, $T_1$ and $T_2$ coherence times, and gate durations) is recalibrated by IBM on an approximately daily basis. To support both real-time inspection and historical analysis, I-QMapper organizes calibration access around a local JSON cache. Upon connection, the tool queries the IBM API and indexes the returned data by its \texttt{last\_update\_date} field. When the user requests calibration data for a previous time point (via the Snapshot, Intraday, or Multi-day viewing modes) the corresponding records are added to the cache, and on every subsequent request the cache is consulted before any new API call is issued. All timestamps are stored in UTC and converted to the user's local timezone only for display, ensuring consistent behavior across geographic locations. The JSON-based cache format was chosen for its human readability and ease of debugging.

\section{QPU Design and General Features}
\label{sec:design}
\begin{figure}[H]
    \centering
    \includegraphics[width=1\linewidth]{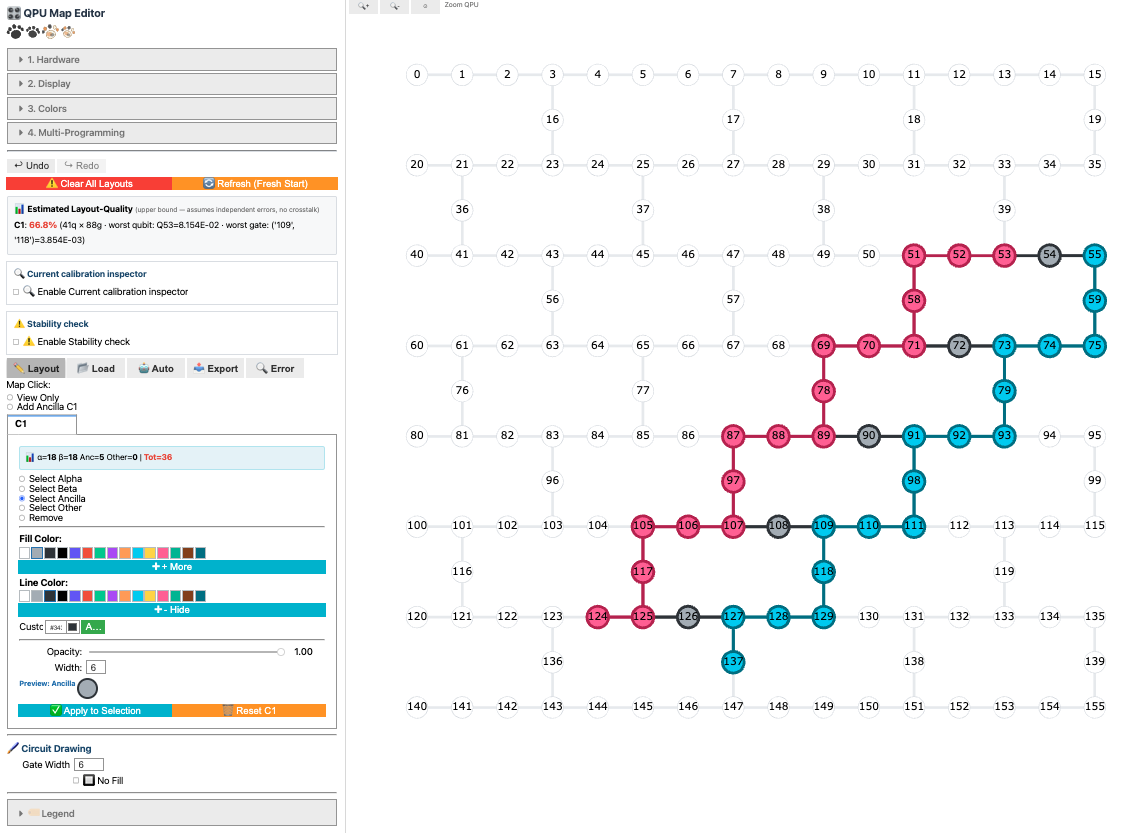}
\caption{Design Mode interface of I-QMapper for the IBM Heron 156-qubit processor (\texttt{ibm\_boston}) with an auto-generated LUCJ layout. Left: control sidebar with four collapsible sections (Hardware, Display, Colors, Multi-Programming), a Layout-Quality Score panel reporting the scoring along with the worst-performing qubit and two-qubit gate, and a tab bar (Layout, Load, Auto, Export, Error) for switching between sub-panels. The Layout sub-panel for the circuit includes role selectors ($\alpha$, $\beta$, ancilla, other), per-role styling controls (fill color, outline color, opacity, line width), and a live counter of assigned qubits. Right: heavy-hex coupling map rendered with native hardware coordinates; each circle is a qubit labeled by its physical index, and edges represent two-qubit gate connectivity.}
    \label{fig:Figure2}
\end{figure}

Design Mode is the default view of I-QMapper and serves as an interactive canvas for manual qubit assignment and circuit construction on the device topology (Figure~\ref{fig:Figure2}). Each qubit is positioned according to its native hardware coordinates, and two-qubit gates are displayed as edges between connected qubits. The processor family is identified automatically from the backend metadata, and the corresponding coordinate layout is selected from a set of topologies defined internally in I-QMapper. The 156-qubit heavy-hex Heron and the 120-qubit Nighthawk topologies are currently supported. Moreover, additional processor topologies can be easily added via the \texttt{constants} module.

The sidebar panel on the left organizes QPU visualization settings into four collapsible sections: \textit{Hardware}, \textit{Display}, \textit{Colors}, and \textit{Multi-Programming}. The first three cover backend selection, rendering dimensions, and general styling options of the QPU representation. \textit{Multi-Programming} operates in two modalities. In \textit{intra-QPU}, multiple sub-circuits are co-mapped onto a single device, each with its own control sub-panel. In inter-QPU mode, two devices are displayed side-by-side, supporting cross-backend layout design and comparison.

The intra-QPU paradigm has been recently demonstrated for LUCJ ansätze on IBM Heron hardware, with parallel and serial executions producing comparable energy estimates after SQD/ext-SQD post-processing~\cite{lucjparallel}.

The \textit{Layout} sub-panel handles manual construction and visual styling. The \textit{General Ansatz} mode assigns selected qubits to a generic role, providing a workflow for circuits without explicit structural constraints. On the other hand, the \textit{LUCJ Ansatz} mode differentiates qubits by their role in the layout ($\alpha$ and $\beta$ spin-orbitals, and ancillary qubits)~\cite{Motta2023LUCJ}. Visual styling (fill color, opacity, outline color and width) is configurable per role, and a live counter reports the number of assigned qubits. All layout edits are fully reversible through dedicated \textit{Undo} and \textit{Redo} controls. A \textit{Clear All} button resets all circuits at once, and a \textit{Refresh} button reloads the device graph and calibration data. For the LUCJ workflow, an automatic layout generator is also available, described in Section \nameref{sec:lqs}.

The \textit{Legend} panel controls the figure legend. Each selected qubit "role" ($\alpha$, $\beta$, ancillary, other, and unselected) has its own checkbox and an editable label. An entry appears only when its role is populated, and entry colors follow the current role styling.

The \textit{Load} sub-panel accepts layouts in \texttt{.json} and \texttt{.npy} formats, as well as qubit lists pasted into a text area. Loading a previously saved session file (\texttt{.qpusession.json}) restores the full layout, style, and calibration reference.

The \textit{Export} sub-panel displays a live JSON preview of the current selection, with qubits grouped either by role or merged into a single list. The layout can be saved to a \texttt{JSON} file, a high-resolution \texttt{PNG} rendering of the device map can be generated via Kaleido for paper-ready figures, and time-lapse exports of the calibration data evolution are available in \texttt{GIF} or \texttt{MP4} format (see Section~\nameref{sec:errors}). Two additional formats support full reproducibility: \texttt{.qpusession.json} stores the complete session state (layouts, calibration timestamp, panel configuration), while \texttt{.qpustyle.json} stores only the visual styling, allowing consistent appearance across sessions.

\subsection{Layout-Quality Score}
\label{sec:lqs}
A persistent banner above the tab bar reports a Layout-Quality Score (LQS) for the active mapping (Figure~\ref{fig:Figure2}). The banner updates automatically as the user adds or removes qubits, either through the \textit{Layout} sub-panel or the \textit{Auto} generator. The LQS is intended as a structural quality metric for the selected set of physical qubits, aggregating the calibration data of the qubits and of the two-qubit gates of the selection. We define:
\begin{equation}
\text{LQS} = \prod_{q \in Q_{\text{layout}}} \left(1 - \varepsilon_q^{\text{ro}}\right) \cdot \prod_{(q_i,q_j) \in E_{\text{layout}}} \left(1 - \varepsilon_{(q_i,q_j)}^{\text{2Q}}\right)
\label{eq:lqs}
\end{equation}
where $Q_{\text{layout}}$ is the set of selected physical qubits (alpha, beta, and ancillas), $E_{\text{layout}}$ is the set of coupling-map edges whose both endpoints lie in $Q_{\text{layout}}$, $\varepsilon_q^{\text{ro}}$ is the readout assignment error of qubit $q$, and $\varepsilon_{(q_i,q_j)}^{\text{2Q}}$ is the two-qubit gate error on the corresponding edge. Each selected qubit and each such edge contributes once, under an independent-error assumption. Single-qubit gate errors, coherence-time effects, and gate repetitions are not included.

The LQS is conceptually related to the cost function defined by mapomatic~\cite{nation2023suppressing}, in that both aggregate per-qubit and per-edge calibration data into a single ranking quantity. The key difference is that our metric is computed directly on the layout and not on the compiled circuit as mapomatic does. Therefore, it provides an estimate of how good the selected set of physical qubits is, based on their calibration data.

The banner also reports the weakest qubit and edge in the layout, i.e. the selected qubit/edge with the highest error, as a quick indication of the layout's limiting element.

\subsection{Calibration Inspector and Stability Check}
\label{sec:calstabcheck}
The \textit{Current Calibration Inspector} exposes the calibration data behind the current layout selection. Dedicated dropdown menus for the qubits (annotated with their role) and for the two-qubit gates list each element from the best to the worst current calibration metric. The enumeration is multi-programming aware: when several circuits are placed, entries from all circuits are listed and disambiguated by circuit label.

The \textit{Stability Check} considers the multi-day calibration history, with a 14-day window fetched by default upon backend connection. For every qubit in the layout, the daily readout error is compared against its median over the selected time window. A day is counted as bad when the error exceeds twice the median, and a qubit is flagged as unstable when this occurs on at least two days. Flagged qubits (and gates, via their CZ error) are listed in a dropdown, ranked by a volatility score $(\max/\mathrm{median} - 1)$, together with the value and date of their worst day. If no element is flagged, the panel reports how many qubits and gates were checked.

\subsection{Automatic Layout Optimization}
\label{sec:autolucj}

The \textit{Auto} engine generates LUCJ layouts in an error-aware fashion. Our implementation builds on the code released by IBM as part of the SQD tutorial for automatic LUCJ layout generation~\cite{ibm_sqd_tutorial}. The original code provides the core graph-construction and subgraph-isomorphism routines: two parallel qubit chains representing the $\alpha$ and $\beta$ orbitals are connected by ancillary qubits, and all valid physical embeddings are enumerated via the VF2 algorithm as implemented in \texttt{rustworkx}~\cite{rustworkx}. Each candidate is scored using a lightweight error-sum function that adds the two-qubit gate errors of all couplers in the layout to the readout errors of all qubits, with an additional coherence penalty for qubits whose $T_1$ or $T_2$ falls below configurable thresholds.

In the original implementation, only the lowest-error layout is returned and the number of ancilla bridges between the chains is fixed at the maximum admissible value. I-QMapper extends this behavior in three ways. First, the full set of valid mappings is retained and exposed in a ranked dropdown menu, so the user can browse candidates, immediately visualize each placement on the device map, and compare their scores. Second, layouts that differ only by an exchange of the $\alpha$ and $\beta$ chain assignments produce identical physical circuits and are de-duplicated. Third, the number of ancilla bridges can be overridden by the user, allowing zero to the maximum number of ancillas, so the user is not restricted to the zig-zag topology.

\begin{figure}[H]
\centering
\includegraphics[width=\textwidth]{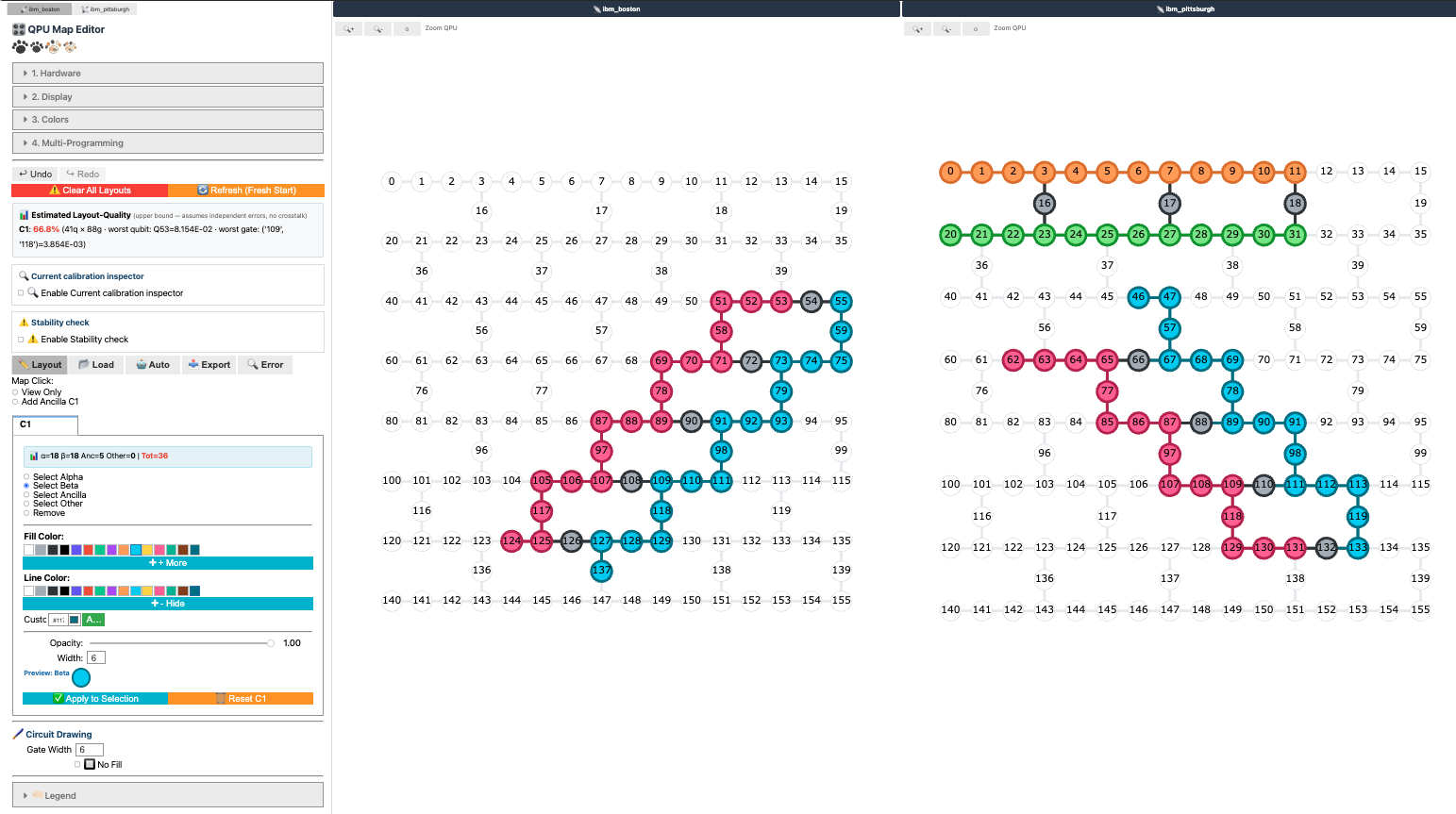}
\caption{Side-by-side comparison of two IBM Heron r3 processors, 
\texttt{ibm\_boston} and \texttt{ibm\_pittsburgh}. \texttt{ibm\_boston} 
hosts a single LUCJ circuit, while \texttt{ibm\_pittsburgh} runs two LUCJ 
circuits simultaneously under the intra-QPU multi-programming regime. 
Distinct calibration profiles drive the auto-engine to different optimal 
placements on the two devices.}
\label{Figure5}
\end{figure}

A further extension supports the \textit{Auto} engine within the intra-QPU multi-programming framework. Given a list of circuit sizes and a user-defined buffer parameter $b$ (the number of graph hops separating distinct circuits), the engine places the circuits sequentially on the residual coupling graph, excluding at each step the qubits already assigned to previous circuits together with all qubits within $b$ hops of them. The buffer prevents adjacent circuits from sharing the same qubits, mitigating potential crosstalk; $b=0$ corresponds to packed placement with no isolation between circuits. When the user manually changes the layout selection of a given circuit, a cascading re-optimization is triggered: all downstream circuits are automatically re-placed on the updated residual graph, keeping the multi-circuit allocation globally consistent. Figure~\ref{Figure5} shows an example of the \textit{Auto} engine applied to LUCJ circuits on two distinct Heron backends loaded in inter-QPU mode: a single LUCJ circuit on the first device, and two LUCJ circuits in intra-QPU multi-programming mode on the second.

The lightweight score inherited from the IBM tutorial and the Layout-Quality Score defined in Section \nameref{sec:lqs} serve different purposes within \textit{Auto}. The error-sum score is the native ranking function of the underlying enumeration routine and is displayed in the ranked dropdown to support the comparison of candidate layouts. The LQS is reported separately in the top banner after a layout has been applied, providing a fidelity-like quantity in $[0,1]$ that is directly interpretable as the quality of the selected layout.

\section{Error Mode and Analysis}
\label{sec:errors}
This panel features two tabs, \textit{Calibration} and \textit{Analysis}, offering calibration visualization together with diagnostic and statistical tools that operate on both current and historical calibration data.

\subsection{Calibration}
Upon switching to \textit{Error Mode}, the QPU topology is rendered as a heatmap with an IBM Quantum-inspired color scale. Several built-in color scale alternatives are available. The two dropdown menus in the \textit{Property \& Scale} section control which calibration property is displayed for qubits and gates separately (Table \ref{tab:calibration-metrics}). Errors units are reported accordingly to the metric utilized by IBM. When a circuit layout is active, the assigned qubits retain their circuit-specific outline color while the fill reflects the calibration data, allowing direct visual assessment of whether the selected qubits fall on high- or low-quality regions of the device. The hover tooltip displays the qubit index, the property name and error value.

\begin{figure}[H]
    \centering
    \includegraphics[width=1\linewidth]{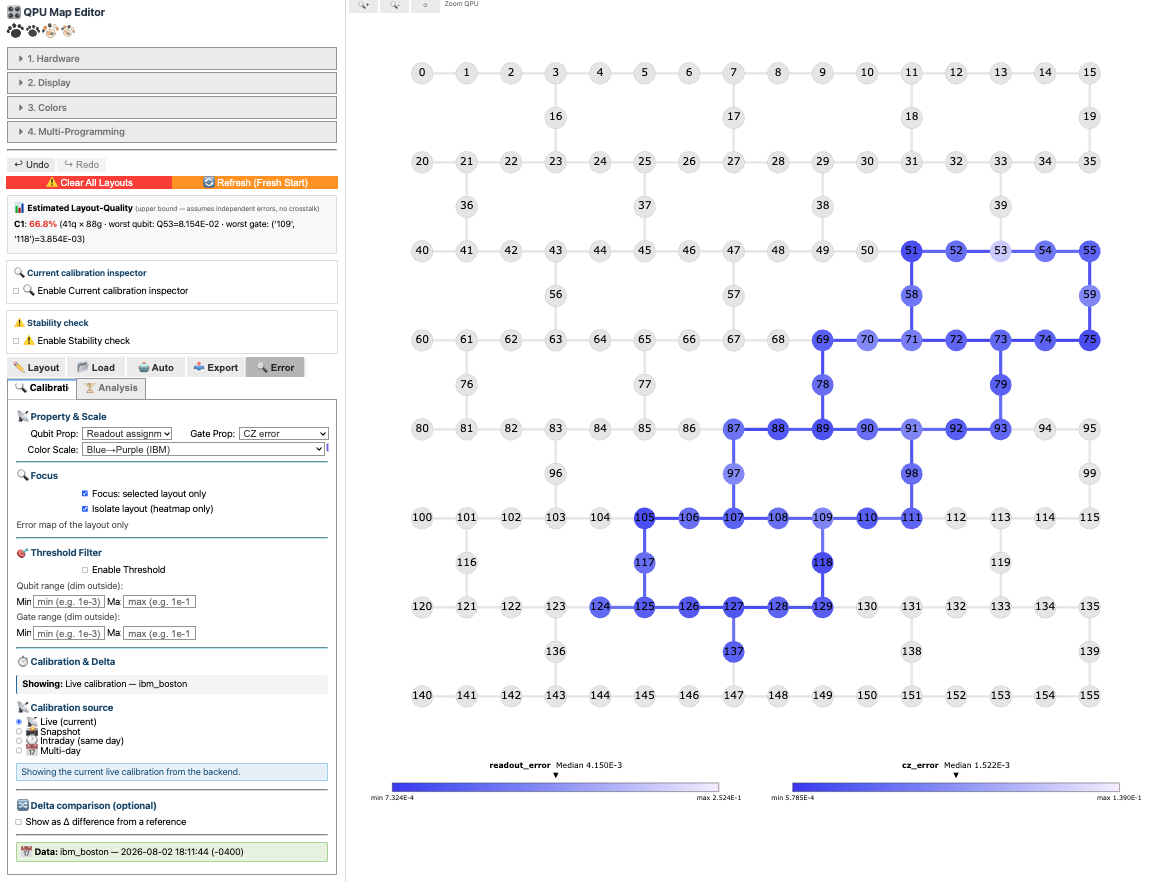}
\caption{The QPU panel in \textit{isolate layout} mode. The device is greyed out and only the layout qubits are filled with the calibration heatmap, exposing the error landscape of the chosen embedding.}
\label{fig:Figure3}
\end{figure}

The \textit{Focus} section restricts the view to the placed layout. \textit{Focus: selected layout only} option greys out every qubit and gate outside the current selection while the assigned qubits keep their circuit styling. \textit{Isolate layout (heatmap only)} instead fills the selected qubits with the calibration heatmap leaving the pure error landscape of the chosen embedding on an otherwise greyed device, Figure \ref{fig:Figure3}. 
\begin{table}[H]
\centering
\begin{tabular}{ll}
\toprule
\textbf{Element} & \textbf{Available properties} \\
\midrule
Qubit & readout assignment error, $T_1$ ($\mu$s), $T_2$ ($\mu$s), \\
      & $P(\mathrm{meas}\ 0 \mid \mathrm{prep}\ 1)$, $P(\mathrm{meas}\ 1 \mid \mathrm{prep}\ 0)$, \\
      & readout length (ns), single-qubit gate errors (ID, RX, $\sqrt{X}$, RZ, X), \\
      & single-qubit gate length (ns), MEASURE error, MEASURE\_2 error \\
Gate  & CZ error, RZZ error, two-qubit gate length (ns) \\
\bottomrule
\end{tabular}
\caption{Calibration properties supported by I-QMapper for qubit-level and gate-level visualization. All metrics are extracted from the standard Qiskit \texttt{backend.properties()} interface. Properties not exposed by a given backend are rendered as N/A.}
\label{tab:calibration-metrics}
\end{table}

The \textit{Threshold Filter} restricts the heatmap to property values within the user-specified range $[\tau_{\min}, \tau_{\max}]$, where $\tau_{\min}$ and $\tau_{\max}$ are the lower and upper bound, respectively. Each element $i$ (a qubit or a two-qubit gate) with property value $v_i^{(p)}$ is rendered in gray and excluded from the active color scale whenever:
\begin{equation}
v_i^{(p)} \notin [\tau_{\min}, \tau_{\max}].
\end{equation}
Thresholds for qubits and gates are set independently, enabling both simultaneous and single-property filtering. This filtering is compatible with all calibration time modes supported by I-QMapper, and summarized in Table \ref{tab:time-modes}. 

The \emph{Live} mode loads the most recent calibration available at connection time. The \emph{Snapshot} option retrieves a specific historical calibration. The \textit{IntraDay} mode probes all 24 hours of a selected day and populates a dropdown menu with the timestamps of the available re-calibrations. Finally, \textit{Multi-Day} operates analogously across a multi-day interval, defaulting to the latest available calibration of each day. The last two modes support time-lapse export in GIF or MP4 format. Both \textit{IntraDay} and \textit{Multi-Day} allow to navigate through the temporal snapshots through an animated slider. 

\begin{table}[H]
\centering
\begin{tabular}{lll}
\hline
\textbf{Mode} & \textbf{Coverage} & \textbf{Navigation} \\
\hline
Live     & connection-time snapshot                  & --- \\
Snapshot & specific date and hour                    & Scan Day dropdown (optional) \\
Intraday & all updates in a 24h window               & animated slider, GIF export\\
Multi-day& latest update per day, in date range interval& animated slider, GIF export \\
\hline
\end{tabular}
\caption{Calibration time-acquisition modes. Threshold filtering and Delta Analysis are compatible with all four modes.}
\label{tab:time-modes}
\end{table}

\textit{Delta Analysis} enables direct element-wise comparison between two calibration snapshots. For each property $p$ and element index $i$, the difference is computed as:
\begin{equation}
    \Delta_i^{(p)} = v_i^{(p),\,\text{current}} - v_i^{(p),\,\text{ref}}
\end{equation}
with the property being either a qubit or a two-qubit gate. The resulting map encodes the change with a diverging color scale: blue for maximum improvement, gray corresponding to negligible change, and red being the highest degradation. When combined with the \textit{IntraDay} or \textit{Multi-Day} slider, each snapshot is dynamically compared against the chosen reference, generating a time-lapse delta visualization that exposes calibration drift and identifies the property whose performance is unstable across the selected time span. Beyond spatial visualization, I-QMapper provides a complementary set of diagnostic tools, described in the following.

\subsection{Analysis}
\begin{figure}[H]
\centering
\includegraphics[width=0.7\linewidth]{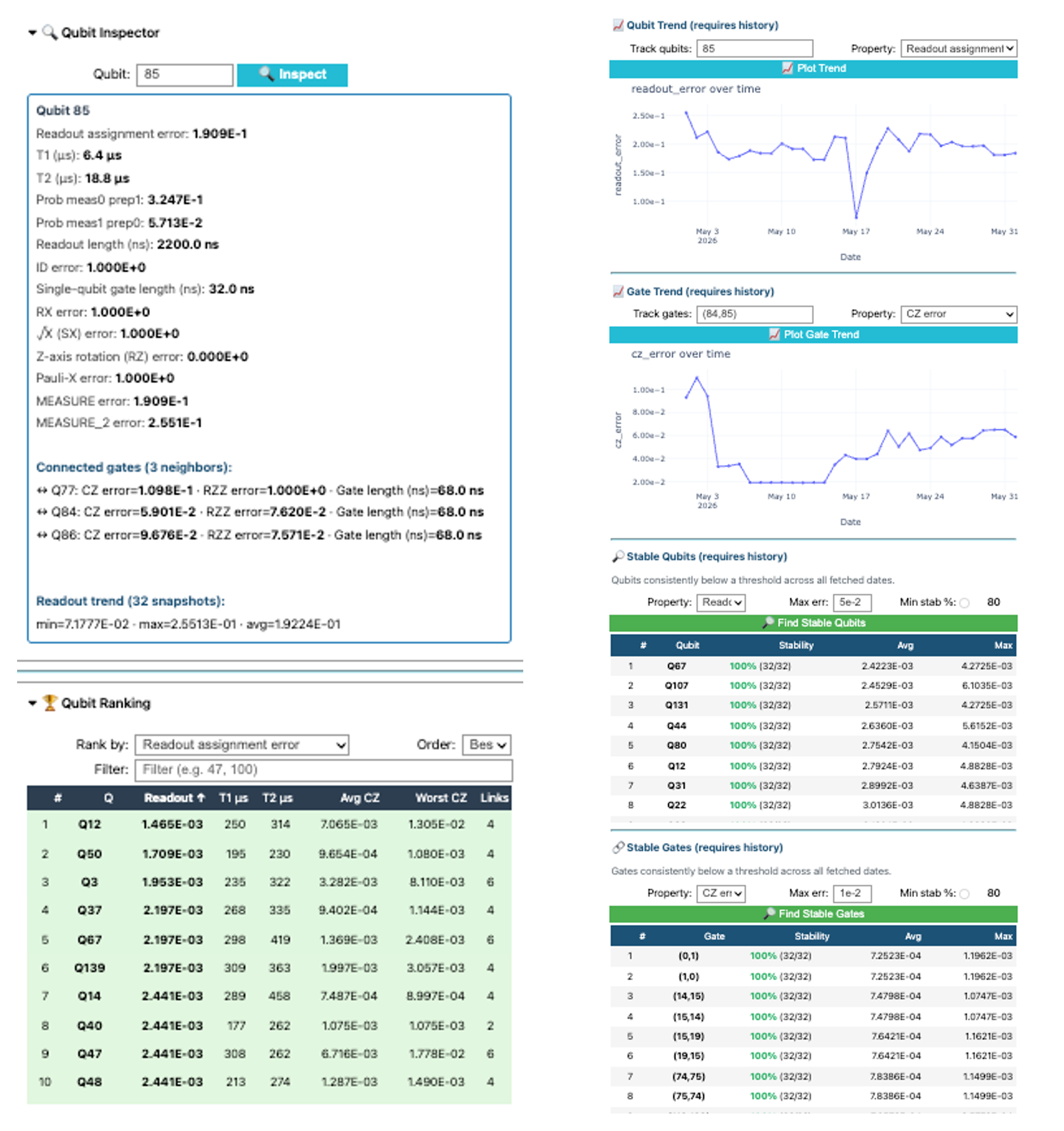}
\caption{The four diagnostic tools of the \textit{Analysis} tab on \texttt{ibm\_boston}. Top-left: the \textit{Qubit Inspector} summarizes all calibration properties of a selected qubit (here Q85) and of its connected gates. Top-right: \textit{Trend} charts show the time evolution of a property across the fetched history, for qubits and gate edges respectively. Bottom-left: the \textit{Ranking} table sorts qubits by any selected property, with auxiliary columns for coherence times and CZ error statistics. Bottom-right: the \textit{Stable Finder} returns qubits and gates whose property stayed within a user-defined threshold for at least a chosen fraction of snapshots.}
\label{fig:Figure4}
\end{figure}

This tab enables the user to access different qubit and two-qubit gate properties for analytical exploration, as shown in Figure \ref{fig:Figure4}. The \textit{Qubit Inspector} displays a summary table of the calibration properties of a selected qubit together with those of its connected gates. The \textit{Ranking} tool sorts qubits and gates by any available calibration property reported in Table~\ref{tab:calibration-metrics}. These properties can also be plotted as a time evolution, via the \textit{Trend} tool, across the \textit{Intraday} or \textit{Multi-day} snapshots. Finally, the \textit{Stable Finder} identifies the elements that maintained the best performance across the calibration history. The user specifies three parameters: the property to evaluate, a threshold value, and a minimum stability ratio (the fraction of snapshots in which the element must satisfy the threshold). For example, setting readout error $< 10^{-2}$ with a stability ratio of $80\%$ returns all qubits that satisfied this condition in at least $80\%$ of the fetched snapshots. Results are sorted by stability, with the most reliable elements listed first. Both \textit{Trend} and \textit{Stable Finder} are particularly useful for experiments spanning multiple days: they identify unstable elements, expose temporal drifts, and let the user prioritize historically reliable qubits over instantaneously good ones, reducing the risk of degraded performance across sessions.

\section{Conclusions and Future Work}

We have presented I-QMapper, an interactive Jupyter-based tool that combines error-aware layout design with calibration analytics on near-term quantum hardware. The error modules expose qubit- and gate-level metrics, supporting Snapshot, Intraday, and Multi-day temporal analyses, threshold filtering, and delta-mode comparison. For quantum-chemistry applications, we implement a dedicated interface for the LUCJ ansatz. Layouts can be generated either manually or automatically, with automatic ones ranked by an error-sum score and reported alongside a complementary Layout-Quality Score. This workflow is compatible with the multi-programming paradigm and enables side-by-side QPU comparison.

Future developments will extend multi-vendor support beyond the current IBM Quantum platform, improve the automatic LUCJ generation routine, and integrate real-time backend characterization.

\section{Code Availability}
The source code of I-QMapper is publicly available at \url{https://github.com/MIQuLab/i-qmapper} under the Apache-2.0 license. The tool can be installed directly from the Python Package Index via \texttt{pip install i-qmapper}; version 0.0.5 corresponds to the release described in this work.

\begin{acknowledgement}

The authors gratefully acknowledge financial support from the National Science Foundation (NSF) through CSSI Frameworks Grant OAC-2209717 and from the National Institutes of Health (Grant Numbers GM130641). 
The authors thank Gurinder Singh and Fabio Cumbo (Cleveland Clinic), together with their IBM collaborators Ella Fejer, Abdullah Ash Saki, Cristina Sanz, Antonio Córcoles, Jessie Yu, Drew DiStefano, and Andrei Constantinescu, for their support and valuable feedback during the development of the code. The authors also gratefully acknowledge Fabio Cumbo, Gurinder Singh, and Abdullah Ash Saki for independently reviewing and validating the implementation.
\end{acknowledgement}

\subsection{LLM Statement}
The conceptualization, scientific design, algorithm selection and development, and architectural decisions are solely the authors' work. The authors provided the domain expertise in quantum computing, including Qiskit integration, calibration data interpretation, and LUCJ ansatz layout strategy. Claude Opus 4.8 (Anthropic) contributed to code implementation, user interface development, debugging, and modular refactoring under continuous author correction, direction, and review. The final code review was performed with Claude Fable 5. Claude Sonnet 4.5 was used to refine the flow and the prose of the manuscript.
\bibliography{AFE_QPU}

\begin{center}
\end{center}
\end{document}